\documentclass[onecollarge,natbib]{svjour2}
\bibpunct{[}{]}{;}{n}{}{,} 
\smartqed  
\usepackage{graphicx}
\usepackage{hyperref}
\usepackage{amsmath}
\usepackage[dvips]{epsfig}
\newcommand{\beq}{\begin{equation}}
\newcommand{\eeq}{\end{equation}} 
\newcommand{\beqa}{\begin{eqnarray}}
\newcommand{\eeqa}{\end{eqnarray}}
\newcommand{\ba}{\begin{array}}
\newcommand{\ea}{\end{array}}

\journalname{Few-Body Systems}
\begin{document}

\title{Quenched dynamics of the momentum distribution of the unitary Bose gas}

\author{Francesco Ancilotto$^{1,2}$
        \and Maurizio Rossi$^1$       
        \and Luca Salasnich$^{1,3}$      
        \and Flavio Toigo$^{1,2}$  }
\institute{$^1$ Dipartimento di Fisica e Astronomia "Galileo Galilei" and CNISM, 
                Universit\`a di Padova, via Marzolo 8, 35122 Padova, Italy 
           \and   
           $^2$ CNR-IOM Democritos, via Bonomea, 265 - 34136 Trieste, Italy
           \and
           $^3$ CNR-INO, via Nello Carrara, 1 - 50019 Sesto Fiorentino, Italy}

\date{Received: date / Accepted: date}

\maketitle

\begin{abstract}
We study the quenched dynamics of the momentum distribution 
of a unitary Bose gas under isotropic 
harmonic confinement 
within a time-dependent density functional approach based on our recently 
calculated Monte Carlo (MC) bulk equation of state.
In our calculations the inter-atomic s-wave scattering 
length of the trapped bosons is suddenly 
increased to a very large value and the real-time evolution of the system is studied.
Prompted by the very recent experimental data of $^{85}$Rb atoms at 
unitarity [{\em Nature Phys.} {\bf 10}, 116 (2014)] we focus on the momentum distribution 
as a function of time. 
Our results suggest that at low momenta, a quasi-stationary momentum distribution is reached 
after a long transient, contrary to what found experimentally for large momenta which equilibrate 
on a time scale shorter than the one for three body losses. 

\keywords{Unitary Bose Gas \and Time Dependent Density Functional \and
          Local Density approximation \and Momentum Distribution}
\end{abstract}

\section{Introduction}

Recent experiments claim to have achieved a metastable 
degenerate gas of ultracold and dilute bosonic 
atoms with infinite s-wave scattering length \cite{liho,brem,zora,corn}.
This is the so-called unitary Bose gas, which is characterized 
by remarkably simple universal laws 
arising from scale invariance \cite{cast}.
While the unitary Fermi gas has been largely investigated 
both experimentally and theoretically 
\cite{book}, its bosonic counterpart has been only marginally theoretically addressed 
\cite{peth,sal1,japs,lee2,sto1,sto2,noi1} because generally 
considered as experimentally inaccessible 
\cite{hoho} due to the dominant three-body losses at very 
large values of the scattering length.

By using a Monte Carlo (MC) approach \cite{noi1} we have 
recently investigated the zero-temperature 
properties of a dilute homogeneous Bose gas by tuning the 
interaction strength of the two-body potential 
to achieve arbitrary positive values of the s-wave 
scattering length $a_s$, while avoiding the formation 
of the self-bound clusters present in the ground-state \cite{noi1}. 
More recently, the MC equation of state has been the key ingredient of a  
local density approximation (LDA) \cite{noi2} to the energy density 
functional to be used in density functional theory (DFT).
Remarkably, the density profiles of a unitary Bose gas in a harmonic trap 
calculated with  DFT using such functional compare very well with the MC ones.
In addition, by using a time-dependent formulation, we have also investigated 
the excitation frequencies as a function of the scattering length. Interestingly,  
the calculated values for the monopole breathing mode, which reproduce 
the expected limiting values of the ideal and unitary regimes,
exhibit a non-monotonous behavior as $a_s$ varies \cite{noi2}. 

Unfortunately, the experimental data on the unitary 
Bose gas \cite{liho,brem,zora,corn} to compare with are
quite scarce, and they do not provide an easy 
tool for validating theoretical approaches.
The most suitable data in this sense are 
the time-resolved measurements  of the momentum 
distribution $n({\bf k},t)$ of a Bose-condensed 
gas quenched at unitarity 
(sudden increase of $a_s$ to a very large value) 
provided by the very recent 
experiment of Makotyn {\it et al.} \cite{corn}.
The main result coming from Ref.~\cite{corn} is 
that, for $k \ge k_B=(6 \pi ^2 n)^{1/3}$ where $n$ is the atom number density, 
$n({\bf k},t)$ evolves to a quasi-steady-state 
distribution on a time scale shorter than the 
one characterizing three-body losses.
We thus try to reproduce the observed phenomenon by solving a nonlinear 
Schr\"odinger equation (NLSE) obtained from 
time-dependent DFT (TDDFT) with our LDA functional, 
based on the MC equation of state and 
where we have introduced a dissipative term 
to take into account three-body losses, whose 
role is relevant during the quenched 
dynamics \cite{metz}. We must notice however, that we expect our single orbital TDDFT to be 
fully reliable only at low momenta, where 
the collective long-wavelength dynamics dominates. 
Therefore our analysis will be confined to low momenta, in a range not directly comparable 
with the one studied experimentally. The results we find, are therefore complementary to those of 
Makotyn {\it et al.} \cite{corn} and suggest that, as expected, while the system equilibrates 
locally in a short time as found by experiments, it takes a very long time to reach quasi equilibrium 
all over the trap. 

\section{Method} 

On the basis of the density functional theory \cite{kohn}, a reliable energy functional of 
the local density $n({\bf r})$ for an inhomogeneous system of interacting bosons at $T=0$
is given by 
\beq 
 E[n({\bf r})] = \int\left\{\frac{\hbar^2}{2m}\left( \nabla \sqrt{n({\bf r})} \right)^2 
                           + n({\bf r})\ \varepsilon(n({\bf r})) 
                           + n({\bf r})\ U({\bf r}) \right\}\ d^3{\bf r} \; , 
 \label{dft}
\eeq
where the quantum-pressure gradient term takes into account effects due to density variations 
\cite{von}, $\varepsilon(n({\bf r}))$ is the energy per atom of the {\it homogeneous} system 
with density equal to the local density and $U({\bf r})$ describes the external confinement, 
which we assume to be an isotropic harmonic potential $U({\bf r})=\frac{1}{2} m \omega_H^2 r^2$ .  
The values of $\varepsilon(n)$ have been recently fitted to the results of 
a MC calculation \cite{noi1} for a wide range of (positive) values of the 
scattering length $a_s$ characterizing the interparticle interaction.
In the weakly interacting regime 
($x\equiv a_s/r_0\ll  1$, where $r_0=(3/(4\pi n))^{1/3}$ is the 
average distance between bosons) the MC results for $\varepsilon(n)$ are very close to 
$\varepsilon _{\rm LHY}(n)$, the universal Bogoliubov prediction \cite{bogo} 
$\varepsilon_B(n)=\frac{\hbar^2}{2m}(6\pi^2n)^{2/3}$ as 
corrected by Lee, Huang and Yang (LHY) \cite{lhyc}. 
In the strong-coupling regime ($x\gg 1$), instead, MC data reach 
a plateau and, in the unitarity 
limit ($a_s\to \infty$), a finite and positive energy per 
particle is found, $E/N=0.70\ \varepsilon_B(n)$. 
The equation of state of the homogeneous system\cite{noi1} 
from such MC calculation can be well interpolated as:
\beq
 \label{fit}
 \frac{\varepsilon(n)}{\varepsilon_B(n)} = \left\{
  \begin{array}{lcc}
   f_{\rm LHY}(x) + a_3x^3  & {\rm for}               & x<0.3     \\
   c_7x^7+c_6x^6+c_5x^5+c_4x^4+c_3x^3+c_2x^2+c_1x+c_0 & {\rm for} & 0.3<x<0.5 \\
   b_0 + b_1\tanh\left(b_2/x - 1 \right)              & {\rm for} & x>0.5
  \end{array}
 \right.
\eeq
with $a_3 = 0.21$, $b_0=0.45$, $b_1= -0.33$, $b_2=0.54$, $c_0 = 4.75$ , $c_1 = -99.72$, $c_2 = 890.68$, 
$c_3 = -4309.56$, $c_4 = 12268.41$,$c_5 = -20488.00$, $c_6 = 18568.27$ and $c_7 = -7052.20$ \cite{nota}.  
In (\ref{fit}), 
$f_{\rm LHY}(x) = \left(\frac{4}{3\pi^2}\right)^{1/3}
x[1+\frac{128}{15\sqrt{\pi}}\sqrt{\frac{3}{4\pi}}x^{3/2}]$ 
is the LHY correction to the Bogoliubov prediction.
Notice that in the deep weak-coupling regime Eq. (\ref{dft}) reduces to the familiar 
Gross-Pitaevskii density functional \cite{gpeG,gpeP} since 
$\varepsilon(n) = {\cal E}_{\rm GPE}(n,a_s) \equiv 2\pi\hbar^2 a_sn^2/m$.
 
In Ref. \cite{noi2} we have shown that the energy functional (\ref{dft}) is very accurate in 
reproducing the MC static density profiles of the inhomogeneous unitary Bose gas under 
harmonic confinement. 
The dynamics of the system can be obtained by generalizing the energy functional (\ref{dft})
into a density-phase action functional $A[n({\bf r},t),\theta({\bf r},t)]$, 
where $\theta({\bf r},t)$ 
is the phase of a single-valued quantum-mechanical wave 
function representing the macroscopic wave 
function of the Bose-Einstein condensate of the superfluid \cite{leggett}, 
which is related to the
superfluid velocity ${\bf v}$ by the relation 
${\bf v} = (\hbar/m)\nabla \theta$ \cite{landau}.
Such a density-phase action functional is written as  
\beq 
A[n({\bf r},t),\theta({\bf r},t)] = \int \Big\{ T[n({\bf r},t),\theta({\bf r},t)] 
- E[n({\bf r},t)] \Big\} \ dt \; , 
 \label{action}
\eeq
where 
\beq 
 T[n({\bf r},t),\theta({\bf r},t)] = \int \left\{- n({\bf r},t) 
                                     \left( \hbar{\partial \theta({\bf r},t)\over \partial t} 
                                            + {\hbar^2\over 2m} 
\left( \nabla \theta({\bf r},t)\right)^2 \right) 
                                     \right\} \ d^3{\bf r} 
\eeq
is the kinetic Lagrangian of Popov \cite{popov}, and  
$E[n({\bf r},t)]$ is given by Eq. (\ref{dft}) under 
the assumption of a time-dependent local density. 
It is straightforward to derive the equations of superfluid 
hydrodynamics with a quantum-pressure term 
(directly related to the von Weizsacker gradient term of 
Eq. (\ref{dft})) by extremizing the action 
functional (\ref{action}) \cite{kim}. 
Moreover, introducing the wave function 
\beq 
 \Psi({\bf r},t) = \sqrt{n({\bf r},t)} \ e^{i \theta({\bf r},t)}
 \label{wave}
\eeq
the equations of superfluid hydrodynamics can be re-written in 
terms of a nonlinear Schr\"odinger equation 
\beq 
 i\hbar {\partial \Psi({\bf r},t)\over \partial t} = \left[ -{\hbar^2 \nabla^2 \over 2m} + U({\bf r}) + 
                                              {\partial (n\varepsilon)\over \partial n} \right ] 
                                                            \Psi({\bf r},t)  \; ,  
 \label{nlse}
\eeq
which is the Euler-Lagrange equation obtained from the 
action functional (\ref{action}) taking into 
account Eq. (\ref{wave}). 

We stress that Eq. (\ref{nlse}) can be alternatively 
obtained from time dependent density 
functional theory (TDDFT) \cite{kohn,ks1,ks2} in the Local Density 
Approximation  using
a single Kohn-Sham orbital to describe the degenerate boson system \cite{kim,noi2}.
The computational approach based on Eq. (\ref{nlse}) has 
been adopted, for instance, to describe 
superfluid helium \cite{trento,anci}, ultracold bosonic 
atoms \cite{kim,noi2}, and also superfluid fermions 
in the BCS-BEC crossover \cite{kim2,manini}. 

\begin{figure}
 \centering
 \epsfig{file=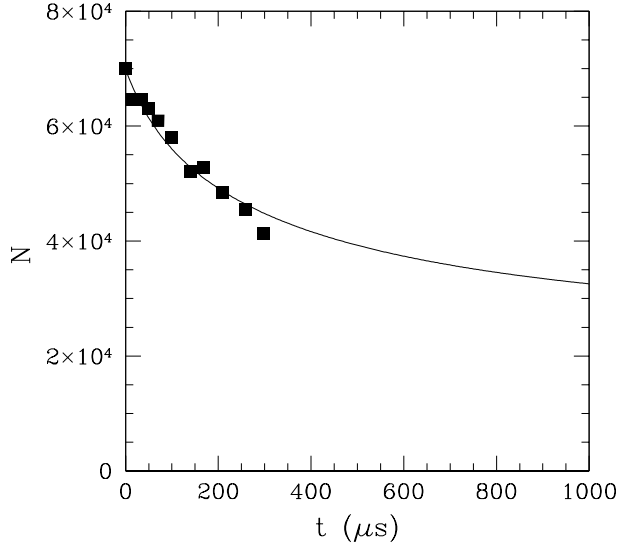,width=10.cm,clip=}
 \caption{Time evolution of the total number $N$ of trapped 
$^{85}$Rb atoms in the unitary Bose gas. 
Solid line: numerical integration of the dissipative nonlinear Schr\"odinger 
equation (\ref{diss-nlse}). 
Filled squares: experimental data from Ref.~\cite{corn}.} 
\label{fig:1}
\end{figure}

In the dynamics of the unitary Bose gas three-body losses play a relevant dissipative role, 
especially close or at unitarity.  
As done in previous applications of Eq. (\ref{nlse}) \cite{metz}, 
we model the effect of three-body losses by adding a phenomenological dissipative 
term $-\frac{  i \hbar L_3 }{2} n^2\Psi$ (with $L_3=9\cdot 10^{-23}$ cm$^6$/sec 
\cite{corn}) in equation (\ref{nlse}).
Thus, the dissipative NLSE we use for our numerical simulations is given by 
\beq 
 i\hbar {\partial \Psi({\bf r},t)\over \partial t} = \left[ -{\hbar^2 \nabla^2 \over 2m} + U({\bf r}) + 
                                                           {\partial (n \varepsilon)\over \partial n} - 
                                          \frac{  i \hbar L_3 }{2}|\Psi({\bf r},t)|^4 \right] 
                                                          \Psi({\bf r},t)  \; . 
 \label{diss-nlse}
\eeq

\begin{figure}
\centering
\epsfig{file=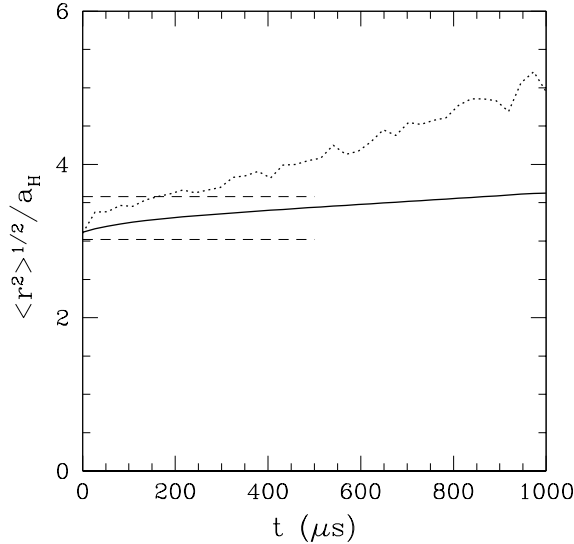,width=10.cm,clip=}
\caption{Scaled average radius $\langle r^2 \rangle^{1/2}/a_H$ of 
the $^{85}$Rb atomic cloud as a function of time $t$. 
Solid line: numerical integration of the dissipative NLSE, Eq. (\ref{diss-nlse}). 
Dotted line: numerical integration of the dissipative GPE, i.e. Eq. (\ref{diss-nlse}) with 
$\varepsilon(n) = {\cal E}_{\rm GPE}(n)$. In both cases the scattering 
length $a_s$ is very large but finite: $a_s=5\cdot 10^5\, a_0$, 
with $a_0=0.53\cdot 10^{-10}$ m the Bohr radius. 
The two horizontal dashed lines give the experimentally-estimated average radius of the 
cloud up to the time $t=500$ $\mu$s \cite{corn}. 
$a_H=\sqrt{\hbar/(m\omega_H)}$ is the characteristic length of the harmonic confinement.}
\label{fig:2}
\end{figure}

We have numerically solved  Eq.(\ref{diss-nlse}) to obtain the real-time evolution closely 
simulating the experimental conditions of Makotyn {\it et al.} \cite{corn}. 
Therefore we consider
as initial configuration  a cloud of $N=70000$ $^{85}Rb$ atoms, 
confined in a spherical harmonic trap 
with frequency $\omega_H=10$ Hz and scattering length 
$a_s = 150\,a_0$ (where $a_0$ is the Bohr radius) 
prepared in its ground state by evolving in imaginary time 
Eq. (\ref{nlse}) without the dissipative term 
(i.e. with $L_3=0$). Then we switch on the dissipative term 
(with $L_3=9\cdot 10^{-23}$ cm$^6$/sec), increase the scattering 
length to a very high, but otherwise arbitrary, value 
$a_s = 5\times 10^5\, a_0$, and let the system evolve in real time for a time $t$. 
At such time we analyze the momentum distribution of the particles, as 
described in the following. 
Notice that according to our calculated 
MC equation of state (\ref{fit}), the chosen value for $a_s$ is
practically equivalent 
to $a_s=+\infty$, i.e. it gives the same energy per atom as
in the truly unitary limit\cite{noi1}.

\section{Numerical results}

Due to the presence of the three-body recombination, whose rate 
increases as $a_s^4$, atoms are
continuously disappearing from the trap. 
At early times, $t \le 350$ $\mu$s, experimental data are 
compatible with an exponential decay  
with a time constant of about $630$ $\mu$s \cite{corn}.
In our simulations the number of particles $N$ in the trap suffers a similar depletion,
driven by the phenomenological dissipative term proportional to $L_3$.
The time-dependent behavior of the total number of trapped atoms can be monitored by solving 
Eq.(\ref{diss-nlse}) and by computing
\beq 
N(t) = \int \left|\Psi({\bf r},t)\right|^2 \ d^3{\bf r} \; .
\eeq
The obtained $N(t)$ is shown as a solid line in Fig.~\ref{fig:1} 
and is compared to the experimental
data of Ref.~\cite{corn}, reported as filled squares. 
Our DFT results are fairly comparable with the experimental points, where these 
are available, but we note that for later times our decay is far from being exponential.
It seems to be characterized by two different time scales:
one, fast, driving the first depletion of the atoms within the trap, 
that covers the experimental data range; 
and a second one, which is much slower, dictating the emptying 
of the trap at very large times. 
We find that, in this second regime, a relevant number of atoms still populate the trap, 
and the depletion is much slower, leading to an 
apparent stationary condition, as discussed in the following, and allowing 
for measurements of properties that give the impression of being equilibrated.

The size of trapped cloud predicted on the basis of our 
DFT calculation is also in reasonable agreement with 
the experimental data, given their uncertainty.
In Fig.~\ref{fig:2} we report the average radius of the 
bosonic cloud as a function of time $t$. 
The solid line refers to the results obtained using the 
dissipative NLSE (\ref{diss-nlse}) with the MC equation 
of state (\ref{fit}) and $a_S=5\times 10^5\, a_0$. 
They are fully compatible with the experimental 
findings \cite{corn} reported as horizontal dashed lines in Fig.~\ref{fig:2}.
For the sake of comparison, in Fig.~\ref{fig:2} we also plot with a dotted 
line the results for the same average 
radius, computed using the dissipative NLSE  Eq. \ref{nlse}
with the same finite scattering length $a_s$ and the 
same three-body loss coefficient $L_3$ 
as above, but with the energy 
density $ \epsilon(n)$ from the Gross-Pitaevskii 
equation of state instead than from MC. 
The figure clearly shows that the radius obtained 
within the GPE theory (dotted line) rapidly exits 
the experimental range as measured in the 
first 500$\mu $s \cite{corn} and displays local 
oscillations due to interference effects in the 
outer part of the atomic cloud  interacting 
with the trap walls. 

\begin{figure}
 \centering
 \epsfig{file=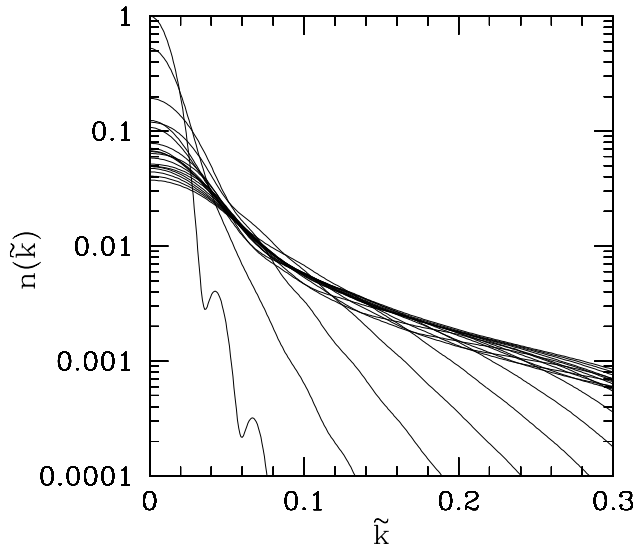,width=10.cm,clip=}
 \caption{Momentum distributions $n(\tilde{k})$ of the bosonic cloud 
of $^{85}$Rb atoms obtained 
at increasing time steps by solving Eq. (\ref{diss-nlse}). 
The narrowest distribution corresponds to $t=0$ 
while the broadest one to $t=1000$ $\mu$s. 
There are $37$ distributions obtained at increasing time steps $\Delta t=27.02$ $\mu$s. 
The data are shown as a function of the reduced wave-vector $\tilde{k}\equiv k/k_B$, where
$k_B=(6\pi ^2n)^{1/3}$ and $n$ is the density at the center of the trap.}
\label{fig:3}
\end{figure}

Given the solution $\Psi ({\bf r},t)$ of the dissipative NLSE 
(\ref{diss-nlse}), the time dependent 
momentum distribution is easily calculated as: 
\beq 
n({\bf k},t) = \left|\int \Psi({\bf r},t) \ \exp{(i{\bf k} \cdot {\bf r})} \ d^3{\bf r}\right|^2 \; ,
\label{nkt}
\eeq

In Fig. \ref{fig:3} we plot $n(\tilde{k},t)$ obtained at different times, 
up to a maximum value of $1$ ms, with $\tilde{k}\equiv k/k_B$, where 
$k_B=(6\pi ^2n)^{1/3}$ and $n$ is the density at the center of the trap. 
Notice that in the figure we report $n(\tilde{k},t)$ at 
low momenta ($0\leq \tilde{k} \leq 0.3$),
because our single-orbital TDDFT is supposed to be fully 
reliable only at long wavelengths. 
During the first $300$ $\mu$s we observe an expansion of the cloud both in real and in 
momentum spaces, and then a quasi-stationary momentum distribution 
seems to develop for larger times.

However, we must point out that at longer times (not shown in the figure) 
$n(\tilde{k},t)$ is still changing. This could be ascribed to the fact that, 
due to the sudden quench of $a_s$ to so high values, the bosonic cloud 
expands until the wave-front interacts with the steeper part of the trap walls and 
is reflected back, letting the cloud to shrink again  at much larger times.
Thus the momentum distribution first decreases (close to the turning point) 
and then it increases again as the contraction of the cloud occurs. 

\begin{figure}
\centering
\epsfig{file=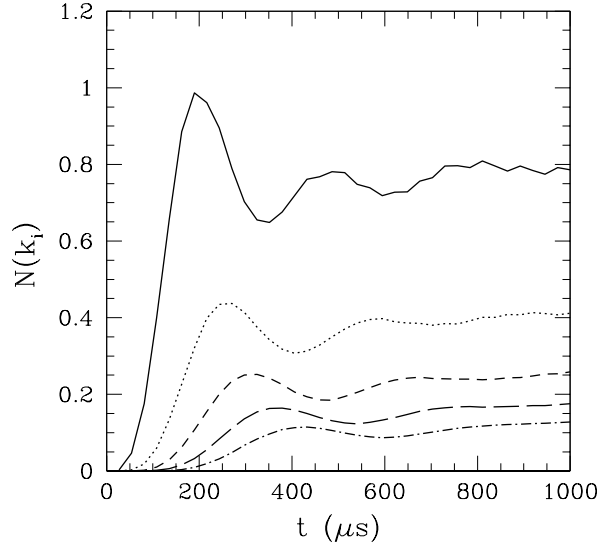,width=10.cm,clip=}
 \caption{Time evolution of the number of atoms $N(k_i)$  with momentum $k_i$
(expressed in units of $k_B=(6\pi ^2n)^{1/3}$). 
The $5$ curves correspond to $5$ values of $k_i$: 
$0.08$ (solid line), $0.12$ (dotted line), 
$0.16$ (dashed line), $0.20$ (long dashed line), 
$0.24$ (dot-dashed line). 
The curves have been arbitrarily normalized to the maximum of $N(k_i=0.08)$ (solid line).}
 \label{fig:4}
\end{figure}

To better clarify the time scales involved in the momentum changes, 
we plot in Fig. \ref{fig:4} the time
evolution of the number of atoms $N(k_i)$ with a 
given momentum $k_i$ for different values of $k_i$ during
the first $\sim 1\,ms$ after the sudden quench. 
We note that the evolution of the distribution for these small momenta 
is not uniform, being oscillatory first, with
the amplitude of the oscillations extinguishing 
on a time scale of $700-800\, \mu s$.

This is to contrasted with what found 
in the experiments of Ref.\cite{corn} for 
 higher ($k> k_B$) components in the momentum distribution.
There, a shorter time scale governs the saturation of the
k-component, increasing from about $100\, \mu s$ to about $300-400\, \mu s$ as $k$ decreases  
from  $k/k_B\sim 1.2-1.3$ to $k/k_B\sim 0.8$.
Our results complement these observations, 
in finding a still larger time-scale when $k/k_B\sim 0.1-0.3$.
Our results can be compared also with recent 
studies on the dynamics of $n(\tilde{k},t)$ after a 
sudden quench to very large value of $a_s$ 
\cite{rancon}. In particular, calculations based on a 
bath approach \cite{rancon} lead to a multistep equilibration process,
where larger $k$ modes equilibrate faster (having a shorter 
relaxation time), the oscillations in $n({\bf k})$ 
are damped on intermediate times, leading to an apparent 
equilibrated state, and the full equilibration is 
reached only on a much larger time scale, when also 
the condensate attains its final time-independent state.

\section{Conclusions}

We have numerically investigated the quenched dynamics 
of a Bose gas of $^{85}$Rb atoms under isotropic 
harmonic confinement after the sudden increase of the 
s-wave scattering length from $a_s=150$ $a_0$ to 
$5\cdot 10^5$ $a_0$. To take into account three body 
losses we have introduced a dissipative term into 
the time dependent nonlinear Schr\"odinger equation 
(TDNLSE) obtained from a local approximation of a 
TD energy functional.
We have shown that, while the equation obtained by using 
the Gross-Pitaevskii equation of state in the 
TD energy functional is unable to fully capture the 
experimental quenched dynamics of the atomic cloud,
the corresponding equation derived from a functional 
containing the energy density fitted to our recently 
calculated Monte Carlo bulk equation of state \cite{noi1} 
gives values of the average radius of the cloud 
in fairly good agreement with very recent experimental 
data \cite{corn}. We take this result as supporting 
the use of our MC equation of state in the 
energy density functional. 

The solution of our dissipative TDNLSE shows a fast 
depletion of the condensate at short times, 
$t \le 300$ $\mu$s due to three-body losses, accompanied 
by the relative increase in the population of 
large momenta. Also this feature is in agreement 
with the experimental findings \cite{corn}. 
As shown in Fig.~\ref{fig:4} though, the momentum densities 
at various momenta reach a 'quasi equilibrium' 
distribution, at very large times of about $1000$ $\mu$s.
One must notice however that even the larger momenta reported in our figures are much smaller
than those studied experimentally, corresponding to values for which our TDDFT is questionable. 
We reconcile our results with those reported in Ref.~\cite{corn} by observing that they refer 
to different length scales: while from the experiment a fast equilibration on a short length 
scale has been found, reminiscent of local equilibrium, our calculation shows that a much longer 
time is needed to get quasi-equilibrium on a large length scale. 
The authors of  Ref.~\cite{corn} interpret 
their results (see their fig. 4) 
as a saturation of the value of $n(k)$ at times of the order of 
$200$ $\mu$s, shorter than the time scale for three body losses ($300$ $\mu$s), thus indicating a
 different mechanism as responsible 
for the equilibration of the momentum distribution. 
Such a mechanism could be the interaction between 
the condensed and non condensed fractions 
of boson atoms which is not present in our 
equation, since the dissipative 
term in our equation lets the total number of atoms to 
decrease while leaving the system in a degenerate state. 

The effect of the interaction between the 
condensate and excited atoms on the dynamics of 
momentum distribution of a Bose gas after a rapid quench 
to a very large value of $a_s$ has been 
explicitly addressed at in some recent 
papers \cite{rancon,yin,kain,sykes,kira,corson} and supports 
the existence of different times domains in the 
dynamics of momentum distribution.
In Ref.\cite{rancon}, in particular, they correspond to a fast 
depletion of the condensate followed or accompanied by a slow 
re-adjustment of the momentum distribution and, 
only at very large times, the reaching of a metastable state. 
We will explore the possibility of coupling the wave function obeying our TDNLSE to 
a bath \cite{rancon} describing the small 
noncondensed fraction with the goal 
of better reproducing the experimental data at large momenta and short times. 

The authors are grateful to J. P. Corson for a critical reading of a previous version 
of this report and for useful remarks. The authors acknowledge for partial support the 
Ministero Istruzione Universit\`a Ricerca (PRIN project 2010LLKJBX).


\begin{thebibliography}{99}
\bibitem{nota} The coefficients $c$'s have been determined by ensuring continuity of the 
               interpolating function and of its first three derivatives at $x=0.3$ and 
               $x=0.5$ once the  $a_3$ and  the $b$'s have been fitted to the calculated 
               values. 
               This procedures automatically well reproduces the MC values in the interval 
               $0.3<x<0.5$.
\end{thebibliography}
\end{document}